# The Fabric of Space-time II

## Gravity by Photon Exchange

**Dr Charles Francis**

**Abstract**
This is one of a number of papers in which the metric for space-time is defined on the subatomic level by means of the interchange of photons, and constrained to be consistent with radar. It is shown that the discrete nature of particle interactions necessitates a small inherent delay in the radar method, and hence the resulting geometry is non-Euclidean. Thus the forces of gravity and electromagnetism can each be seen as an effect of the same interaction between elementary particles, and we can identify gravitational and inertial mass.

Dr Charles Francis
Lluest, Neuaddlwyd
Lampeter
Ceredigion
SA48 7RG

# Gravity by Photon Exchange

## 1  The Fabric of Space-time

This is one of a number of papers examining the implications of a theory of matter in which space-time is conceived as a fabric of composed of discrete particle interactions. There have been previous suggestions that at a fundamental level of physical law, variables such as time should be discrete [1], and there are many references in the literature on the potential quantisation of gravity which suggest that this may be so [2]. Just as we recognise that the surface of a sheet of paper is fibrous and non-geometrical, we may also conceive that ontological space is not smooth or continuous. The microscopic structure of a manifold such as that used in General Relativity is described by use of limits, which have no empirical basis, and should be seen in the light of Riemann's mathematical definition [3] which deliberately ignores the question of whether the manifold represents something ontological. Riemann had this to say (Clifford's translation)

Either, therefore, the reality which underlies space must form a discrete manifold, or we must seek the ground of its metric relations outside it.

It appears to us that there is an interpretation of modern physical law which does provide the ground for metric relations which are not a reflection of an ontological manifold underlying space, but rather of relationships found in particle interactions. The motivation for this is that quantum electrodynamics has shown that the exchange of photons is responsible for the electromagnetic force, and so for all the structures of matter in our macroscopic environment. But as Bondi pointed out, the exchange of photons is also the process used to measure space-time co-ordinate systems [5] by the radar method. It is not unreasonable, therefore, to postulate that photon exchange generates all the geometrical relationships in the macroscopic environment, just as it generates these relationships in the results of measurement by radar.

To imagine the substructure of space-time we conceive that each charged particle follows some repeating process according to which we may regard a primitive notion of time as one of its ontological properties. We call this notion a time-line. For example a particle may have the possibility of emitting or absorbing a photon in each discrete instant of its time-line. This can be considered a repeating process adequate for the notion of primitive time. Whenever an exchange of photons takes place, i.e. a photon is emitted by one charged particle and absorbed by another, and a photon is then immediately emitted by the second particle and absorbed by the first, then a coordinate for the second particle is established in terms of the time-line of the first.

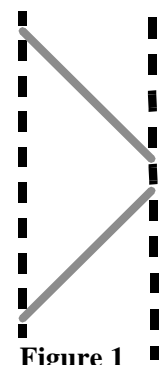

**Figure 1**

Since visualisation involves the awareness of geometry, it is clear that the pre-geometric properties of matter cannot strictly be visualised. Nonetheless, a "stitch in space-time" can be illustrated diagrammatically (figure 1). Charged particles are shown as dashed lines, where the dashes represent discrete intervals of time in the particle's time line. Photon exchange is shown by continuous grey lines. A single stitch such as that shown can only give a single value of distance and time for the second particle. Many stitches are required to determine the properties of space-time. It is not necessary to assume that all these stitches require the immediate return of a photon, only that they use photon exchange and combine to a consistent geometry.

It is legitimate to imagine a system of such particles in the form of a diagram provided that it is understood that the space between the lines of the diagram has no theoretical or ontological meaning. The properties of space-time depend on the internal relationships between lines and nodes in the diagram, not on the geometrical properties of a drawing. In systems of many particles such as we generally observe, photons are constantly exchanged, and macroscopic space-time can be construed as some kind of composition or average of the primitive space-time associated with photon exchange.

Obviously we cannot empirically analyse an individual exchange of photons, since observation would disrupt it. But we can statistically analyse the effect of many such exchanges. Macroscopic behaviour is predicted from the statistical analysis of the behaviour of many particles. Since the process of photon exchange is the same as we use in radar, the average behaviour of a system in which there are many such exchanges should obey the geometrical relationships found empirically. In radar the time and distance of an event are defined by a process in which a photon is emitted and one returns. The theory states the same geometrical relationships are generated whenever photons are emitted and photons return. As this process is taking place all the time in the subatomic structure of matter geometrical relationships are found almost everywhere.

Since the macroscopic reference frame constitutes some form of average of the behaviour of individual particles and each individual particle is in part responsible for the generation of the macroscopic reference frame, individual particles do not in general have exact position and uncertainty in position becomes a feature of the description of elementary particles. Mathematical uncertainty is the subject of many valued logic [7]. In many valued logic the position of a particle may be described by a truth function representing a measure of certainty that it will be found at each position. The truth function gives an out of focus (fuzzy) image of a point-like particle viewed from a macroscopic reference frame.

In a typical measurement in quantum mechanics we study a particle in near isolation. The implication is that there are too few ontological relationships to generate classical properties, such as the property of position. Then the property of position does not exist prior to the measurement, and the measurement itself is responsible for introducing physical relationships sufficient to generate the property of position. In this case we cannot describe the state of matter in terms of position, but we can associate with it a truth function which has no direct ontological analogue but which can be used for calculating the probability of a given result of measurement. Quantum position is defined as the truth function for a measurement of position. Weighted logical OR can be then defined for quantum position using the laws of vector space, and states of definite momentum can be defined by trigonometric series [8]. Thus the principle of superposition is seen as a definitional truism deriving from a naming convention for states of matter in which there is no precise definition of position.

The interpretation that the wave function in quantum mechanics is a set of truth values in a many valued logic is consistent the orthodox interpretation as expressed by Dirac

The expression that an observable 'has a particular value' for a particular state is permissible in quantum mechanics in the special case when a measurement of the observable is certain to lead to the particular value, so that the state is an eigenstate of the observable...In the general case we cannot speak of an observable having a value for a particular state, but we can...speak of the probability of its having a specified value for the state, meaning the probability of this specified value being obtained when one makes a measurement of the observable.

But we do not assume that a measurement need be done to put the state into an eigenstate. A configuration of matter can give rise to a definite value for a measurable property whenever the net behaviour of the particles in that configuration generates that property, whether or not that property is observed or measured. Thus states of matter may be eigenstates of observables even when no observation has been carried out.

I reviewed the derivation of special relativity from the radar method in [9]. Although it is not easy to see how to derive the four dimensional properties of Minkowsky space-time or spin directly from photon exchange between charged particles, it is possible to take the argument the other way. In [10] I have shown that the empirical properties of the measurement of space-time require an underlying substructure of matter in which charged particles emit and absorb photons. Newton's laws and Maxwell's equations were then also be derived for the same structure. The derivation shows that quantum position can be embedded into the wave function, and rests heavily on the uniqueness of the Dirac equation [11]. It strongly suggests that the reason for a four dimensional universe with spin is that this is the smallest, and possibly the only, number of dimensions in which a solution exists.

In [9] and [10] I assumed that the laws of physics are formulated for inertial reference frames, given by the definition: *Any two observers in inertial reference frames will each measure equal and constant red shift in observations on the other's reference frame*. The purpose of the present paper is to show that the discrete nature of particle interactions implies a modification must be made to the theory. It will be seen that the geometrical relationships defined by photon exchange do not permit the definition of perfectly inertial reference frames. The consequence will be the non-Euclidean property necessary to generate the force of gravity.

## 2 Non-Euclidean Geometry

When Euclid wrote down the axioms of geometry he was aware that neither he nor his contemporaries had been able to prove his fifth postulate, (that parallel lines can extend indefinitely always the same distance apart), but for two thousand years it was subjected to attempted proof. Newtonian space is rooted in Euclidean geometry, and as a metaphysical notion has been repeatedly and severely challenged by philosophers and mathematicians. Gauss expressed severe doubts about the a priori truth of Euclidean geometry and was perhaps the first to believe that it was not necessarily true. He organised expeditions into the Alps to measure whether the sum of the angles of a large triangle is $180^o$, but did not detect a deviation from Euclidean geometry.

If the geometrical properties of matter are regarded as empirical properties of particle interactions then there is no reason to believe in Euclid's fifth postulate. If we measure the distance AB between any two points, and measure two equal distances AD and BC perpendicular to AB, as in figure 2, then we have no prior reason to assume that AB is equal to CD, as measured by an observer at C. Of course there is nothing to stop us from constructing a Euclidean co-ordinate system, in which the distance CD is defined to be the same as AB. It is often convenient to do so, but we do not assume that distances calculated in a Euclidean co-ordinate system are the same as distances found by direct measurement. We also cannot assume that it is meaningful to extend a Euclidean co-ordinate system indefinitely in all directions out into space; coordinates are only meaningful in regions of the universe where it is possible to (directly or indirectly) measure distances.

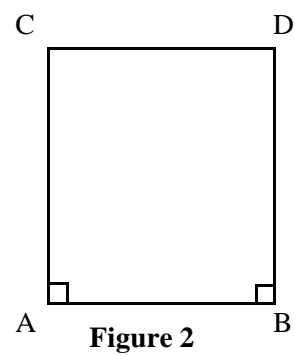

**Figure 2**

Suppose that A and B are close together so that one clock at A can be used to measure (i.e. define) the time for both A and B, and one clock at C measures the time for both C and D. Now suppose that in a region of space the geometry is such that the longer AD and BC are, the shorter the distance CD becomes, as measured by radar based on the clock at C. A physicist at A defines a Euclidean reference frame, in which CD is a fixed length equal to AB. Then it appears to the physicist that radar based on the clock at C gives a shorter distance for CD because the clock at C is running slow. By the principle of uniformity the same is true of any physical process. So the frequency of the wave function of a particle moving from B to C also gets slower, and the wavelength gets longer, i.e. it is red shifted. The logical truth function for position is embedded in the wave function, so it is natural to ask what effect red shift will have on measurements of position.

Place ABDC between Cartesian axes, O$x$ and O$y$ as in figure 3. Let CD be related to AB by a linear formula

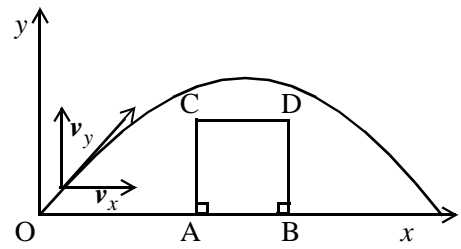

2.1 $\qquad CD = AB(1 - gy)$

where $y$ is the height of CD above AB, and $g$ is a constant. Consider a ball thrown up with initial velocity $v_0$. Let the velocity during the flight of the ball be $v$, with components $v_x$ and $v_y$. Motion in the $x$-direction causes no change in geometry, and no red shift, but motion in the $y$-direction causes the wave function for each particle in the ball to be red shifted by the law that the frequency of the wave function is proportional to the speed of a clock at height $y$. Energy is frequency by definition [10], so the motion of the ball is described by a truth function for measurements of position satisfying

**Figure 3**

2.2 $\qquad E = E_0(1 - gy)$

The total energy of the ball is the sum of the expectations of the energies of the particles in the ball, each one subject to the same red shift. Thus, using linearity to reinterpret $E$ as this expectation, 2.2 becomes the classical energy equation for the ball in a uniform gravitational field.

### 3    The Relationship between Energy and Curvature

It is only necessary to calculate the gravity due to the curvature in space-time caused by a single elementary particle. If there is other matter in the vicinity it also generates red shift. The effect of all such red shifts is linear. It will be seen below that the geometry of space-time has a singularity of total curvature proportional to mass-energy at each elementary particle, and is Lorentzian elsewhere. In a macroscopic reference from this does not appear as a singularity, as there is uncertainty in position. Quantum position is then embedded into a wave function defined on a differentiable manifold with curvature proportional to mass-energy density. The linearity of tensors is then sufficient to state Einstein's field equation for general relativity [4] (without the cosmological term).

A photon is emitted by particle **A**, and absorbed by **B**, and one is reflected back to **A** (figure 4). There is an inherent delay in the reflection due to the discrete nature of particle interactions. It is not obvious that there is a single interval of discrete time between the absorption and emission of the photon but there must be a characteristic lag of magnitude $T$ for the reflection to take place (the inherent lag in return of radar is a feature of the interaction between elementary particles, and applies to macroscopic phenomena only as the expectation of many elementary particle interactions). The lag can be written as a vector in **B**'s reference frame, and is therefore proportional to the energy momentum vector in that frame. Thus, for some constant $G = G_\mathbf{B}$

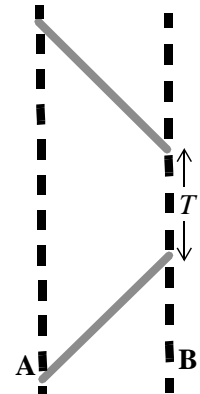

**Figure 4**

3.1 $\qquad T = (T, \mathbf{0}) = G(m_\mathbf{B}, \mathbf{0})$

After Lorentz transformation this appears in **A**'s reference frame as

3.2 $\qquad T = G(M, \mathbf{P})$

where $M$ is the mass-energy and $\mathbf{P}$ is the momentum of **B**.

Recall that the quantities of time and distance are defined by algebraic formulae:

3.3 $\qquad \text{time} = \dfrac{(t_1 + t_2)}{2} \quad \text{and} \quad \text{distance} = \dfrac{(t_1 - t_2)}{2}$

Time and distance may be mapped onto a space-time diagram. If this were an accurate representation, then the co-ordinate system could be perfectly mapped onto the Euclidean geometry of a flat piece of paper. But the reflection of light is really two events, the absorption and the emission of a photon. These two events have the same space-time coordinate. We cannot redefine geometry by simply subtracting an additional distance $GM$ to the distance calculated from radar for three reasons. First, the value of time would still be poorly defined at the event. Second, we have no knowledge of what the value of $GM$ actually is; it may have a different value depending on the type of particle that the photon reflects off. Third, subtracting $GM$ does not permit a consistent definition of geometry; before we subtract $GM$ we must know what red shift to apply, and before we know red shift we must have defined geometry.

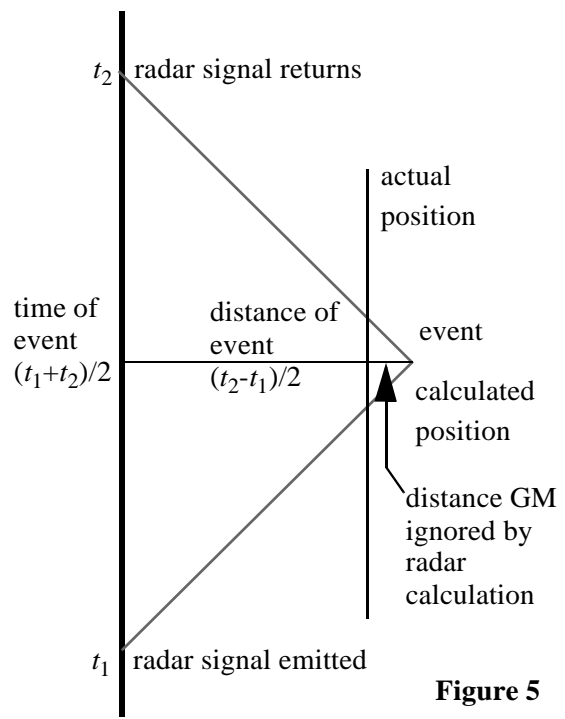

**Figure 5**

We can calculate the effect on the geometry of space-time by comparing the distance, *r*, of an event as measured by radar, with the distance, *s*, along a path from the event to the clock calculated from the metric defined by adding together small distances along the path, each measured individually by radar in that part of the path. Figure 6 is a space-time diagram based on the metric distance *s* along a path from the event to the clock. Then it can be seen that the speed of the clock must be red shifted by a factor *k* so that the event can be plotted at a point. We can determine curvature if we can calculate *k*. By definition

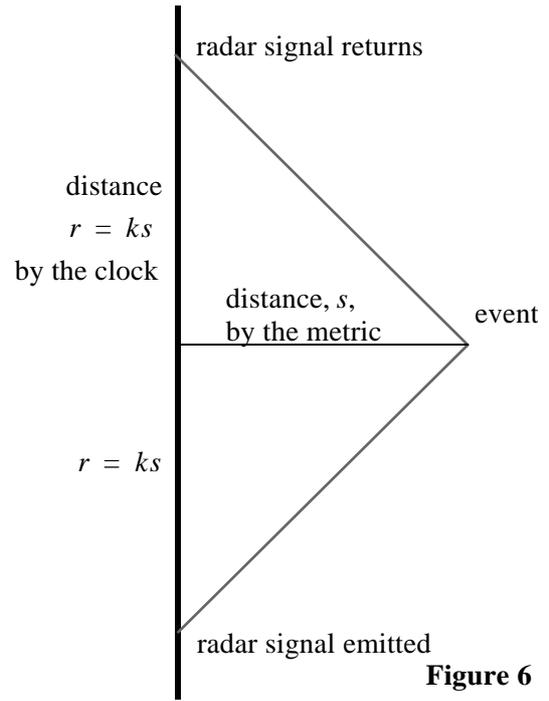

**Figure 6**

3.4 $\qquad r = ks$

Differentiating,

3.5 $\qquad dr = s\,dk + k\,ds$

Because *s* is the sum of small distances, *ds*, calculated by direct measurement we have, close to the origin

3.6 $\qquad ds = dr$

Substituting and rearranging the terms,

3.7 $\qquad \dfrac{1}{s}ds = \dfrac{1}{1-k}dk$

Integrating

3.8 $\qquad \ln s = \ln\left(\dfrac{A}{1-k}\right)$

where *A* is a constant to be determined. So

3.9 $\qquad s = \dfrac{A}{1-k}$

and by 3.4 we obtain

3.10 $\qquad r = s - A$

The minimum distance which could be measured by radar is $r = GM$, i.e. the characteristic lag at $s = 0$. Thus $A = -GM$ and we have

3.11 $\qquad r = \left(1 + \dfrac{GM}{s}\right)s$

Comparison with 3.4 gives the formula for red shift

3.12 $\qquad k = \left(1 + \dfrac{GM}{s}\right)$

# Gravity by Photon Exchange                                                                                 7

Apply this red shift to the wave function for every particle in an isolated body of initial energy $E_0$, and take the expectation to find the classical energy equation for a body in a spherical gravitational field.

$$3.13 \qquad E = E_0\left(1 + \frac{GM}{s}\right)$$